\documentclass{rsifpublic}
\usepackage{graphicx}
\usepackage{epsf}
\usepackage{amssymb,amsfonts,amsmath}
\usepackage{xcolor}
\usepackage{epstopdf}
\usepackage{lineno}
\usepackage{soul}
\usepackage{color}
\usepackage{ulem}
\usepackage{hyperref}
\begin{document}
\runningheads{S. Mishra, W. van Rees and L. Mahadevan}{Coordinated  crawling via reinforcement learning}
\begin{topmatter}

 \title{Coordinated crawling via reinforcement learning}

\author{Shruti Mishra$^{1}$, Wim M.\ van Rees $^{2}$, L. Mahadevan$^{1,3}$\corrauth}
\address{ $^{1}$ Paulson School of Engineering and Applied Sciences, Harvard University, Cambridge, MA 02138, USA\\
$^{2}$ Department of Mechanical Engineering, Massachusetts Institute of Technology, Cambridge, Massachusetts 02139, USA\\
$^{3}$Department of Physics, Department of Organismic and Evolutionary Biology, Harvard University, Cambridge, Massachusetts 02138, USA}

\begin{abstract}
Rectilinear crawling locomotion is a primitive and common mode of locomotion in slender, soft-bodied animals. It requires coordinated contractions that propagate along a body that interacts frictionally with its environment. We propose a simple approach to understand how these coordinations arise  in a neuromechanical model of a segmented, soft-bodied crawler via an iterative process that might have both biological antecedents and technological relevance. Using a simple reinforcement learning algorithm, we show that an initial all-to-all neural coupling converges to a simple nearest-neighbor neural wiring that allows the crawler to move forward using a localized wave of contraction that is qualitatively similar to what is observed in {\it D. melanogaster} larvae and used in many biomimetic solutions.  The resulting solution is a function of how we weight gait regularization in the reward, with a tradeoff between speed and robustness to proprioceptive noise.  Overall, our results, which embed the brain-body-environment triad in a learning scheme, has relevance for soft robotics while shedding light on the evolution and development of locomotion.  
\end{abstract}
\keywords{crawling, locomotion, learning, biomimetics}
\end{topmatter}
\corraddr{(lmahadev@g.harvard.edu)}

\section*{Introduction}

 The locomotion of an animal is a result of coordination of its nervous system with its body and environment \cite{chiel1997brain}. Understanding coordinated motions that involve sensory feedback and proprioception requires a theoretical framework integrating the brain, body and environment \cite{rossignol2006dynamic,pehlevan2016integrative}. But how do these smooth rhythmic motions arise in the first place?

%Any complete understanding of locomotion necessarily requires integrating the brain, body and environment into a theoretical framework \cite{rossignol2006dynamic,hoyt1981gait,pehlevan2016integrative}, to understand how coordinated motions can arise using sensory feedback and proprioception and lead to walking, running, swimming etc.   \cite{biewener2018animal,holmes2006dynamics, tytell2011spikes}. But how do these smooth rhythmic motions arise in the first place?

Experiments on locomotory dynamics in model systems, such as the fly larva of {\it D. melanogaster} \cite{heckscher2012characterization}, suggest that early in larval morphogenesis, neurons are part of a well-connected network. During development, the pruning of neuronal connections reduces the connectivity of neurons via both biochemical and biomechanical feedback modulated by behavior and function embodied in twitching that gradually gives way to coordinated locomotion \cite{narayanan2014developmental, berni2012autonomous}. In the larva and more generally in many soft bodied organisms, motion arises via rectilinear crawling \cite{eltringham1971life,trueman1968burrowing}, wherein rhythmic contraction and relaxation of  muscles create waves that propagate either forward (prograde) or backward (retrograde) along the length of the body. This induces forward locomotion when the interaction with the substrate is asymmetric, e.g.\ when friction in the forward and backward direction are very different. The asymmetry in friction has both a passive and an active component: the presence of anisotropic denticles allows the body to slide more easily in one direction than another passively, while dorso-ventral muscles can partially lift the body to modulate friction actively \cite{heckscher2012characterization}. In either case, the result is the conversion of waves of contraction to net motion of the body.

Substantial previous experimental work characterizing {\it D. melanogaster} crawling  has highlighted the role of sensory feedback in initiating and maintaining the gait \cite{suster2002embryonic} and has inspired recent theoretical work  on the dynamics of a segmented, soft-bodied  {crawler} moving on a frictional surface \cite{Paoletti2014,pehlevan2016integrative}. These studies have shown that minimal representations of the musculature and neural dynamics suffice to explain a number of these experimental observations that include the onset and propagation of contractile waves that lead to locomotion, and further suggest that the rhythmic gait can arise without a central pattern generator. Here, neural impulses drive the activation of muscle forces, resulting in  deformation of the body, producing biomechanical strain. Proprioceptive sensing of this strain in turn drives neural impulses, thereby closing the feedback loop. The result is that the crawler moves forward by simultaneously lifting and contracting its body segments, starting from the posterior segments, and moving towards the anterior end. Critically, in these and most other studies, the neural system is assumed to have a fixed, predetermined connectivity. 

Since the muscles, body wall and connective tissue in the body of a {\it D.\ melanogaster} larva develop asynchronously \cite{suster2002embryonic}, a natural question is how these subsystems are wired together for robust performance. Indeed, could the crawler use proprioceptive feedback to  {learn} a coordinated gait for forward crawling, i.e.\ rewire the neuronal connections using experientially driven sensory feedback to achieve a coordinated gait, as observed experimentally \cite{suster2002embryonic}? To explore  this, we use the framework of reinforcement learning (RL) \cite{sutton1998reinforcement}. Originally inspired by observations of how animals learn to perform certain functions, the approach has gained significant traction recently in the context of training computers in games
\cite{silver2017mastering}, strategies for moving through a fluid \cite{colabrese2017flow,reddy2016learning}, and other domains. We frame our question in terms of the coupled dynamics of a neurophysical system for the crawler and a reinforcement learning algorithm for neuronal rewiring, using sensory feedback to maximize a reward associated with crawling forward. 

\begin{figure}[h]
\includegraphics[width=1\columnwidth]{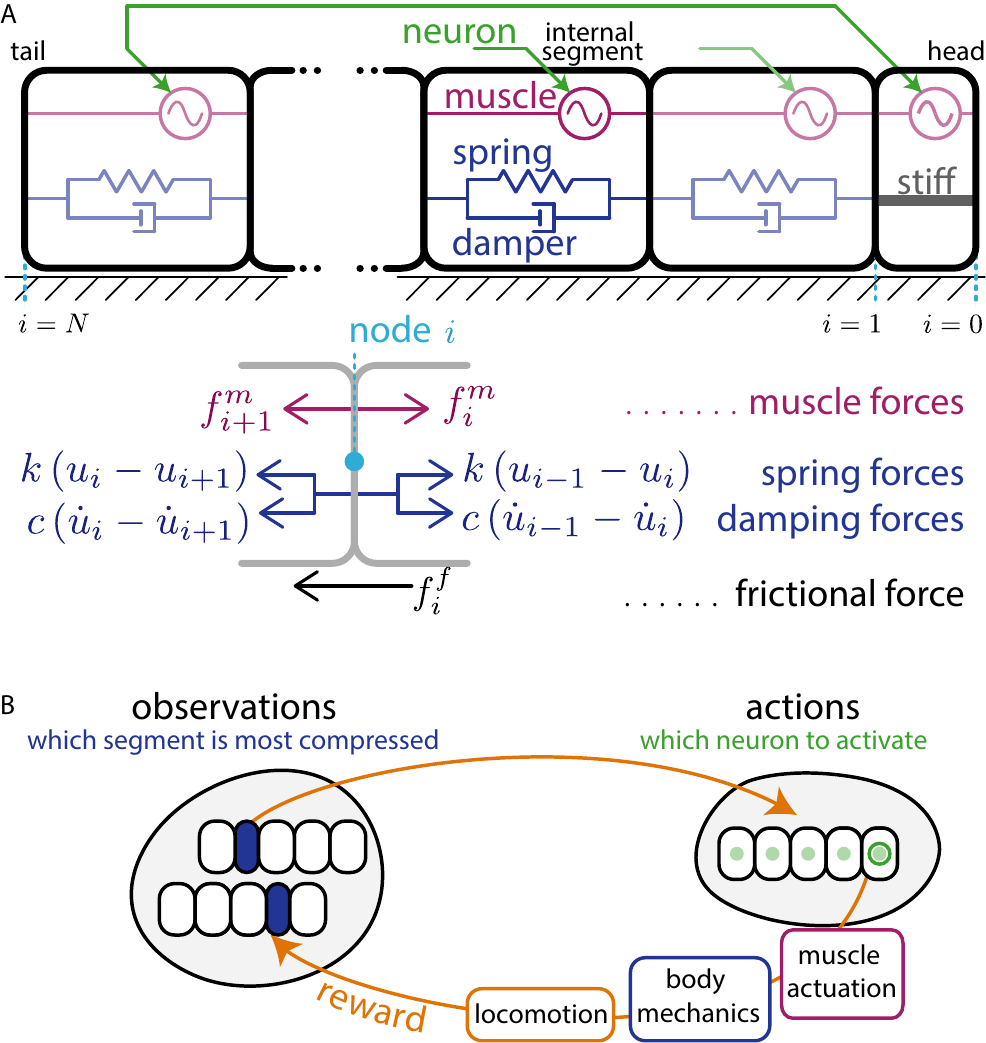}
\caption{Schematic of the crawler. (A) Each segment of the soft-bodied crawler is represented by a spring-damper system and a muscle. Each muscle acts to stretch the segment and is driven by a single neuron. 
(B) Interactions between the different components of the crawler as it learns using the feedback from its environment. }
\label{fig:model}
\end{figure}
%
% \begin{itemize}
% \item damped spring, segments
% \item muscles
% \item neurons
% \item anisotropic friction 
% \end{itemize}
\section*{Mathematical model of crawler}

Our mathematical model is chosen to mimic a soft-bodied crawler, the {\it D. melanogaster}  larva, which has 10 segments connected at their boundaries (nodes), as shown in Figure \ref{fig:model}A \cite{Paoletti2014}. Each segment is assumed to have a passive viscoelastic response, and can be actively contracted by muscles that respond to neuronal inputs as schematized in Figure \ref{fig:model}A. The firing of a segmental neuron causes muscular activation to deform the segment which then moves if the forces overcome friction; simultaneously the segment also transmits forces to neighboring segments where neurons can be activated if the strain crosses a threshold. This leads to a propagating wave even in the absence of a central pattern generator.  We now turn to quantify the three sub-systems corresponding to the body, the brain and the environment. 

\subsection*{Mechanical model}
The segment boundaries, or nodes, $i \in [0,10]$, are mechanically characterized by their displacements $u_i$. All the segments are assumed to have a stiffness $k$, and damping constant, $c$. Each segment deforms due to a contractile force $f_i^m$  exerted by a muscle $i$ and due to a frictional force $f_i^f$ from the external environment at node $i$. Ignoring the role of inertia, since the animals move slowly, force balance at node $i \in [1,9]$ in Figure \ref{fig:model}A implies that
\begin{eqnarray}
k\left(u_{i+1}-2u_i+u_{i-1}\right) 
+ c\left(\dot{u}_{i+1}-2\dot{u}_i+\dot{u}_{i-1}\right) \nonumber \\
+ f^{m}_i - f^{m}_{i+1} = f^{f}_i.
\end{eqnarray}
The force-balance equations at the head and the tail are different from those at the internal nodes as the head and tail do not have a segment ahead of and behind them, respectively. At the head ($i=0$), 
\begin{equation}
k\left(u_{1}-u_0\right) 
+ c\left(\dot{u}_{1}-\dot{u}_0\right)
+ f^{m}_0 - f^{m}_{1} = f^{f}_0,
\end{equation}
while at the tail ($i=N=10$),
\begin{equation}
k\left(u_{N-1}-u_{N}\right) 
+ c\left(\dot{u}_{N-1}-\dot{u}_{N}\right)
+ f^{m}_{N} = f^{f}_N.
\end{equation}
%It is possible to derive a continuum model in the limit of large segment number from the above equation \cite{Paoletti}, but here we limit ourselves to a discrete model to hew closely to the best studied example of a crawler, the {\it D. melanogaster} larva.

\subsection*{Neuromuscular model}
For muscular activity in a segment, we use a model that responds to the timing of neuronal spikes with a built-in temporal decay constant $\tau_m$ and a limiter to set the maximum force amplitude so that
\begin{equation}
\tau_f \frac{df^m_i}{dt} = -f^m_{i}+F_{\max}^m \min\left[1,F_i^m(t)\right],
\end{equation} 
\begin{equation}
F_i^m(t) = \Sigma_{t^s \in \left\{t_i^s\right\}} e^{-(t-t^s)/\tau_m}
\label{eqn:musc-decay}
\end{equation}
For the neuromuscular dynamics, we use the simple $\theta$-model \cite{ermentrout1986} to drive the activation of neuron \emph{i}, where $I_i(t)$ is the time-dependent input to the neuron  \emph{i}, and $\tau_\theta$ is the time-scale of neuronal activity: 
\begin{equation}
\tau_\theta \frac{d\theta_i}{dt} = 1-\cos \theta_i + (1+\cos\theta_i) \min\left[1,I_i(t)\right].
\label{eqn:theta}
\end{equation}
In the $\theta$-model, the neuron `spikes' every time the value of $\theta$ crosses a multiple of $2\pi$, so that the set of spike times $t^s$ for neuron $i$ is given as 
\begin{equation}
\left\{t_i^s\right\} = \left\{t\vert \ \mathrm{mod} \left(\theta_i(t)-\pi,2\pi\right)=0\right\}.
\end{equation}
\subsection*{Environmental friction model}
Finally, for the interaction of the crawler with the environment, we use an asymmetric friction law so that forward motion experiences less friction than backward motion. In our one-dimensional model, this acts as a proxy for both the passive and active components of the friction associated with the structure of the ventral surface and the ability of crawlers to lift up their segments as they crawl forward \cite{heckscher2012characterization}. Furthermore, we impose the condition that the friction force vanishes whenever $\dot{u} = 0$, and require a smooth transition between the positive and negative values for forward and backward velocity, so that the friction force is given by equation \eqref{eqn:friction}, where $\eta_f$ is the ratio of maximum frictional forces in the forward and backward directions, $\epsilon^f$ is a smoothing parameter, and $\dot{u}^0$ is a constant chosen such that $f^f(0)=0$,
\begin{equation}
f^f\left(\dot{u}\right) = 0.5f_{\max}^f\left[\left(1+\eta_f\right)\tanh\left(\frac{\dot{u}-\dot{u}^0}{\varepsilon^f}\right)
+\left(1-\eta_f\right)\right].
\label{eqn:friction}
\end{equation}
All together, our mathematical model eq.\ (1-8) determines the gait and locomotion of the crawler: given the neural connectivity weights and an initial neural impulse leads to an input that drives eq.\ (6) and through this, drives eq.\ (4) and eqns.\ (1-3).

%~~~~~~~~~~%~~~~~~~~~~%~~~~~~~~~~%~~~~~~~~~~%~~~~~~~~~~%
%                          PARAMETER CHOICES				
%~~~~~~~~~~%~~~~~~~~~~%~~~~~~~~~~%~~~~~~~~~~%~~~~~~~~~~%
\subsection*{Scaling and parameter choices}
We scale the relevant variables in our model using the time-scale of neuronal activity $\tau_\theta$, the equilibrium length of a segment $L$ and the stiffness of a segment $k$. Then the dimensionless parameters corresponding to the variables presented in the mechanical model are: $\tau_f/\tau_\theta$- the ratio of timescales for muscular and neuronal activity, $c\tau_\theta/k$ - the dimensionless damping, $f_{\max}^f/kL$ - the scaled maximum frictional force, and  $F_{\max}^m/kL$ - the scaled maximum muscular force. The specific values for these nondimensional parameters used throughout this work, given in Table S1, are consistent with experimental estimates for a {\it D.\ melanogaster} larva \cite{pehlevan2016integrative}.  

For a given gait of the crawler, such as the coordinated gait shown in Figure \ref{fig:learning}A and in supplementary video 1, we can compare our results to those for a {\it D.\ melanogaster} larva using the scaled segment deformation $\Delta u/L$, the characteristic wave speed, $v \ \tau_\theta/L$, and the speed of the larva $v_{\mathrm{crawler}}\ \tau_\theta/L$. For the parameter values from Table S1, the peak contraction of a segment  is 33\%, consistent with experiments \cite{hughes2007sensory}, yielding a wave speed of 0.026 waves/$\tau_\theta$ and a forward speed of 0.0056$L/\tau_\theta$. Using the value of 1.5 waves/s and a length of 4 mm for a third instar larva from \cite{hughes2007sensory}, implies that $\tau_{\theta}=17$ ms and $L = 4/10 = 0.4 \ \mathrm{mm}$, respectively, so that the forward speed of the crawler is $0.13 \ \mathrm{mm/s}$. Using a wave speed of 0.5-1.5 waves/s and a length of 1 mm for first instar larvae, we get a range of $\tau_\theta$ of $17-51 \ \mathrm{ms}$, which translates to a forward speed of $11-33 \ \mu$m/s, compared to the observed range of $45-120 \  \mu\mathrm{m/s}$  \cite{heckscher2012characterization}.

\section*{Reinforcement learning (RL) strategy}
With the established physical model and parameter choices for the crawler, we turn to RL to determine the neural weights for efficient crawling. The framework of RL consists of an  {agent} interacting with its  {environment}, with the aim of achieving a goal. An agent moves through different environmental  {states} by taking  {actions}. As it does so, it accumulates  {rewards} from the environment, with the goal of taking  {actions} that maximize its long-term rewards, itself a discounted sum of successive rewards. This goal is achieved by  {learning} a mapping  that links an {action} to its current environmental  {state}; this mapping is known as the agent's  {policy}. The RL description is summarized in Figure \ref{fig:model}B. 

\subsection*{Formulation of state, action and reward}
In our formulation, the  {observation} of the agent is an incomplete knowledge of itself and its frictional environment. Given the established importance of proprioception \cite{Paoletti2014} in locomotion, it is likely to be  important in the learning process as well. A minimal approach accounting for this is via the observation $o$ associated with the index of the segment that is most strongly contracted, since that requires knowledge of a single variable that can be easily computed via a series of pair-wise comparisons. Then
\begin{equation}
o = \mathrm{argmin}_{i \in \left(1, \dots, N\right)} \left(u_{i}-u_{i-1}\right)
\label{eqn:state}
\end{equation}
The action $a$ is the input to the $\theta-$model that drives neuronal activity, resulting in muscle actuation i.e.\ $I_i(t)$ in equation \eqref{eqn:theta}. We further restrict this by allowing the input $I_i(t)$ to have values of 0 (OFF) or 1 (ON), with only one neuron active at a given time,
\begin{equation}
    I_i(t) = \begin{cases}
    1 & \text{if $i=a$} \\
    0 & \text{if $i\ne a$}
    \end{cases}, \qquad a\in\{0, \dots, N-1\}
\end{equation}
Here, we note that it is possible for several segments to have active muscles even though only one neuron can be active at a particular time, because the muscle forces can decay much more slowly than neural activity, depending on the ratio $\tau_m/\tau_{\theta}$. Noting that experimental observations of larval crawling show that the head and the tail move together \cite{heckscher2012characterization}, we activate the tail neuron $I_N$ every time the head neuron is activated i.e.\ when $I_0 = 1$, we set $I_N = 1$.

Since the goal is to move forward, we set the reward $r$ accordingly,
\begin{equation}
r = \left(\bar{u}_{t+\Delta t} - \bar{u}_t\right) - \epsilon r_2,
\label{eqn:reward}
\end{equation}
where $\bar{u}$ is the position of the centroid of the crawler, $t$ denotes time, $\Delta t$ is the size of the discrete time step (Table S1), and $r_2= \max_i\left(\left|u_{i+1}-2u_i+u_{i-1}\right|\right)$ is a penalty on large variations in strain along the length of the crawler, with $\epsilon$ determining the relative contributions from this strain gradient to the reward $r$.   
%
%\begin{equation}
%r_2 = \max_i\left(\left|u_{i+1}-2u_i+u_{i-1}\right|\right).
%\label{eqn:pen-sg}
%\end{equation}

%~~~~~~~~~~%~~~~~~~~~~%~~~~~~~~~~%~~~~~~~~~~%~~~~~~~~~~%
% ~~~~ LEARNING SUMMARY 
%~~~~~~~~~~%~~~~~~~~~~%~~~~~~~~~~%~~~~~~~~~~%~~~~~~~~~~%
We use a form of RL known as Q-learning \cite{sutton1998reinforcement}, with a discrete representation for the state and action spaces. The entries in the Q-matrix, $Q(s,a)$, represent how much cumulative  {reward} the crawler expects to get after taking an action $a$ in a state $s$, i.e. $\sum_{k=0}^\infty \gamma^k r_{t+(k+1)\Delta t}$, where $\gamma \in [0,1)$ is the discount factor (Table S1) that weighs the long term rewards vs the short term rewards. To maximize the expected discounted cumulative sum of rewards, the entries $Q(o,a)$ are updated each time the agent takes an action in a state, according to the update rule   
\begin{equation}
    Q(o,a) = (1-\alpha) Q(o,a) + 
    \alpha\left(r_t + \gamma \max_a\left(Q(o',a)\right)\right),
\end{equation}
where $\alpha$ is the learning rate, and $o'$ is the subsequent observation made by the agent. The policy is a greedy policy, meaning that in each state, the agent takes the action that corresponds to the highest value. The learning is done in  {episodes}; each episode corresponds to the crawler moving a fixed distance forward, after which it is reset to its original undeformed configuration.  The crawler goes through a number of episodes in this manner, gaining experience in the interactions between neurons, body-mechanics and environment, updating its Q-matrix as it goes through the episodes. It is worth emphasizing that our learning algorithm has just two parameters, a learning rate $\alpha$ and a discount factor $\gamma$, in contrast to many recent variants of RL that have many hyper-parameters; thus most reasonable choices for these will converge and yield similar policies. We choose $\alpha = 0.05$ to allow for stochastic effects and $\gamma= 0.95$ to strive towards the case of high long-time rewards \cite{sutton1998reinforcement}.
 
 \subsection*{Experimental results: regularized and unregularized gaits}
We initialize the crawler in an undeformed state, with a Q-matrix of values that are uniform and high.  Then the crawler is equally likely to take any action independent of the state of the crawler, and since the values are high, i.e. the reward is lower than the expected reward, the crawler explores other actions. This leads to uncoordinated gaits; an example is shown in Figure S1. As the Q-matrix converges towards its steady-state value, the rewards become closer to the expectation of the crawler, and the policy converges. and the  experience of the crawler through subsequent episodes eventually leads to a coordinated gait by means of a converged policy.

\begin{figure}[h]
\includegraphics[width=1\columnwidth]{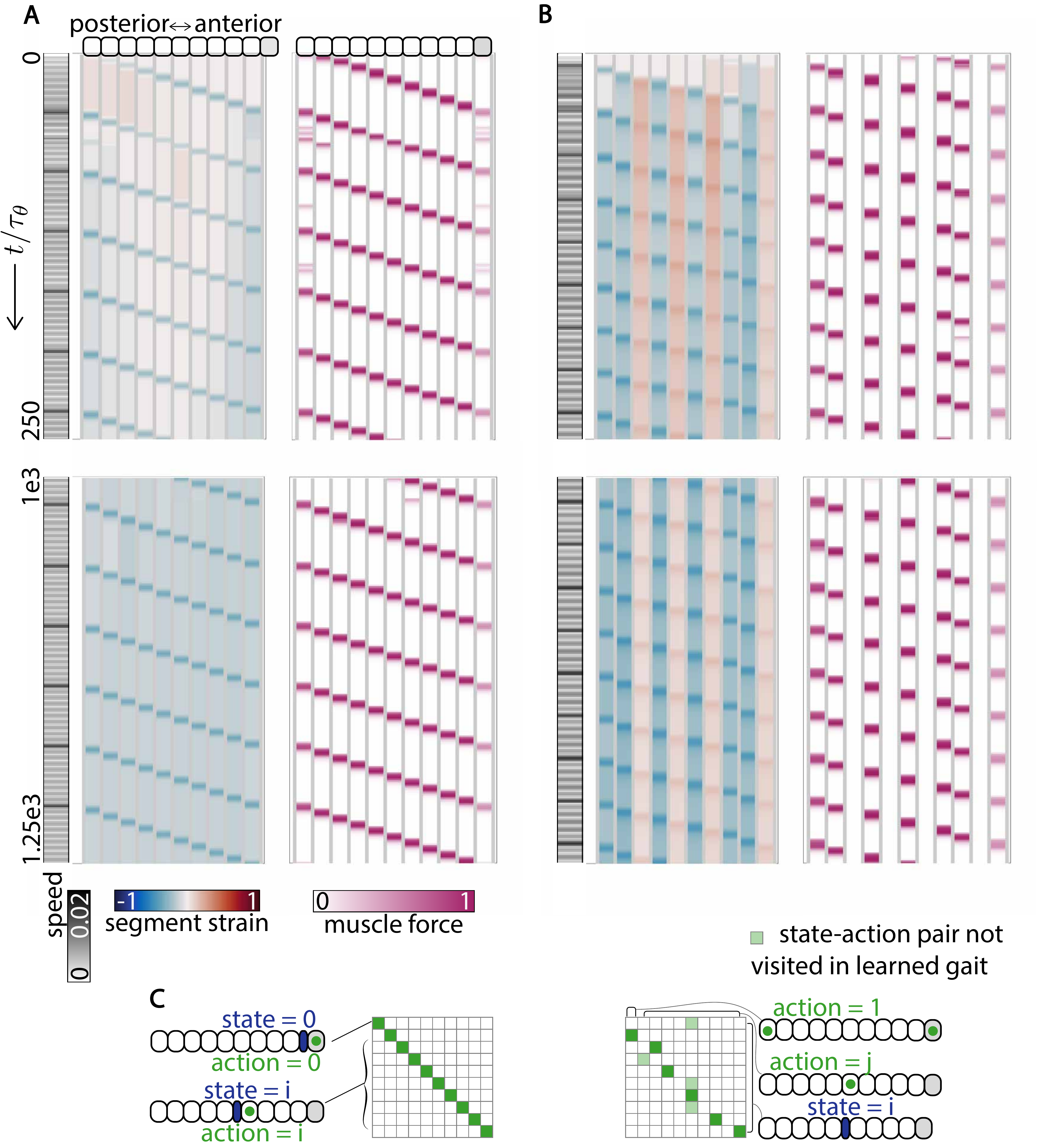}
\caption{Learning of coordinated gaits in a neurophysical model determined using eq.\ (1-12).(A) shows a regularized gait ($\epsilon = 0.01$), with the 10 segment positions and the strains/muscle forces within them   and (B) is an unregularized gait ($\epsilon = 0$).  The parameter values are summarized in Table S1. (C) Converged policy corresponding to the gaits in (A) and (B), with the green and light green squares corresponding to $\pi(a|s)=1$ in the final policy and light green squares corresponding to states which are never reached in the converged gait. }
\label{fig:learning}
\end{figure}

 Figure \ref{fig:learning} shows two coordinated gaits corresponding to two values of the regularization parameter $\epsilon=0.01$ (Fig.\ 2A) and $\epsilon =0$ (Fig.\ 2B), as defined in equation \eqref{eqn:reward}.  In both of the gaits, the crawler moves by means of a traveling wave of contraction from tail to head. The regularized gait corresponds to observations of a   larva consistent with experiments, wherein a localized wave causing sequential segmental contraction moving from tail to head as shown in Figure \ref{fig:learning}A. In contrast, the unregularized gait, corresponding to $\epsilon = 0$ is characterized by a 10\% higher speed, and larger variations in segment strain, and is due to the fact that some muscles are never activated (Figure \ref{fig:learning}B, right), leading to pairs of segments moving together (see SI video 1).  The policies for both gaits are shown in Figure \ref{fig:learning}C. These results justify our use of a regularization penalty in the reward to recover gaits that are biologically plausible and are also consistent with the diagonal neuronal weights that result. 

\begin{figure}[h!]
\includegraphics[width=1\columnwidth]{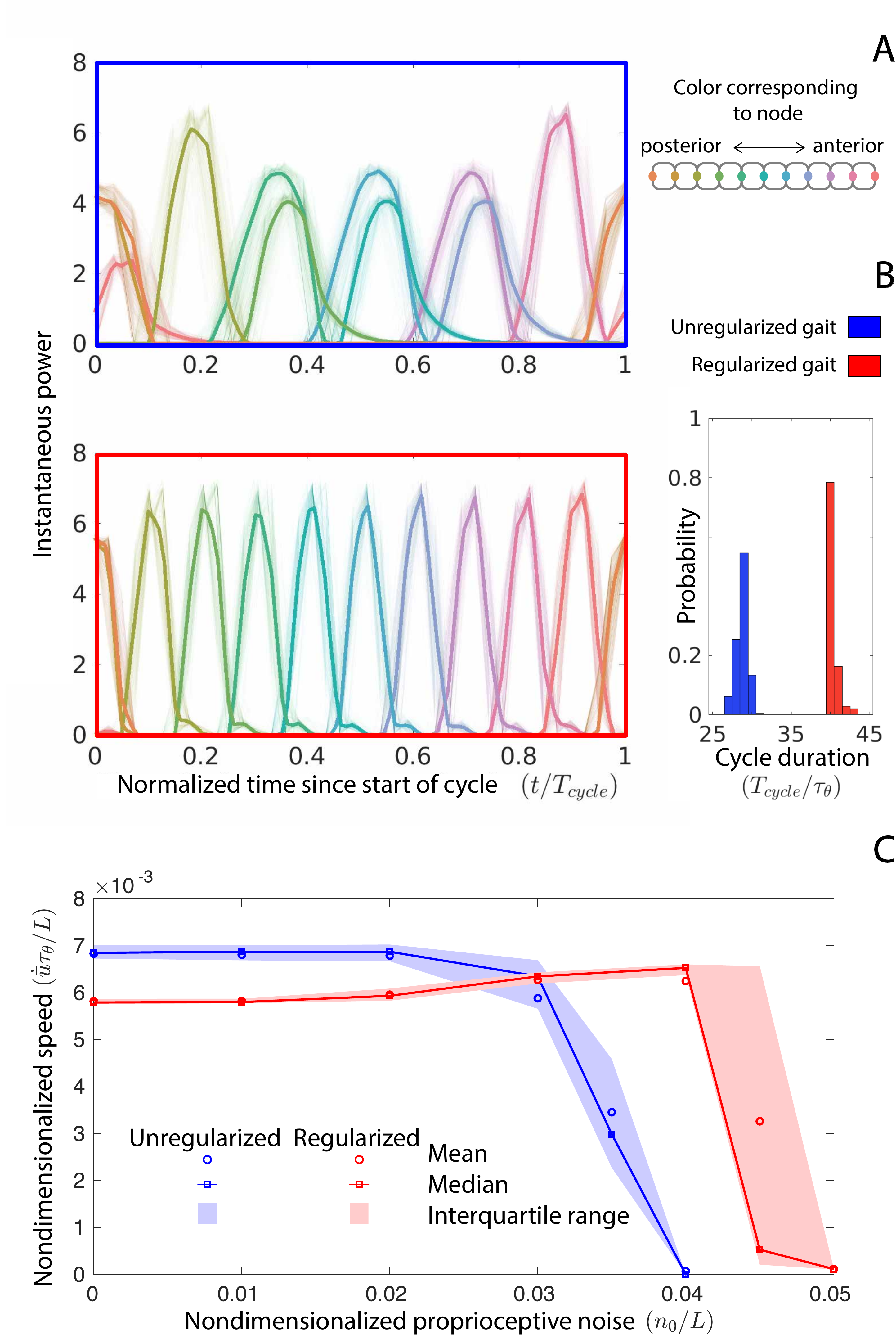}
\caption{Comparison of the unregularized (blue) and regularized (red) gaits. (A) Power as a function of phase in a cycle for each node, for a number of cycles, with the different colors corresponding to the different nodes. (B) Duration of a cycle for the unregularized and regularized gaits. (C) Speed versus proprioception noise for the unregularized and regularized gaits.} 
\label{fig:learning-compare}
\end{figure} 

To further compare the  gaits, we show the power expenditure, cycle duration and robustness to noise in Figure \ref{fig:learning-compare}. The power exerted at each node,
\begin{equation}
    p_i = \left|f_i - f_{i-1}\right| \left|u_i\right|,
\label{eqn:power}
\end{equation}
is a periodic function for both cases. For the regularized gait, the maximum power and the duration for which power is non-zero, are both more uniform across the interior nodes, while for the unregularized gait, there is a larger variation in power across nodes (Figure \ref{fig:learning-compare}A). Figure \ref{fig:learning-compare}B shows the distribution of cycle duration for the two gaits, and shows that the higher speed of the unregularized gait is achieved via a faster propagation of waves along the length of the crawler. 

To test whether these policies are robust, we explored the response of the two gaits to uncertainty in the crawler's ability to sense proprioceptive strain. We implement this by replacing the deterministic observation of the most compressed segment, given by \eqref{eqn:state}), by a noisy version  with $o = \mathrm{argmin}_{i\in(1\dots N)}\left(u_i - u_{i-1} + U \right)$    where $U\in [-s,s]$ is a uniformly distributed random variable and $s$ is the maximum amplitude of the noise. We find that while the regularized gait has a lower speed than the unregularized gait at low levels of noise $s$, as the noise level increases, the regularized gait maintains its speed, while the unregularized gait does not, as seen in the crossover in Figure \ref{fig:learning-compare} (bottom right), showing a tradeoff between speed and robustness to noise. Comparing the segment strain over the course of a cycle, we observe that the unregularized gait varies over a smaller range as compared to the regularized gait (denoted by a smaller contrast in colors for a particular segment in Fig.\ 2B vs Fig.\ 2A). This suggests that the unregularized gait should be more susceptible to proprioceptive noise, consistent with what is observed in Fig.\ 3C. 

\section*{Discussion}
Our minimal approach to learning a coordinated gait in rectilinear crawling embeds the question of determining the neural weights via reinforcement learning in a broader framework linking the brain, the body and the environment and shows that we can recover propagating contractile waves similar to experimental observations \cite{heckscher2012characterization}  and theoretical studies  \ \cite{Paoletti2014,pehlevan2016integrative}. Regularizing the reward to penalize strain gradients provides smooth gaits that expend power more uniformly in space and time, as well as gaits that are robust to uncertainty in the crawler's ability for proprioception, but at the cost of speed. Indeed there is a tradeoff between speed and robustness when these gaits are challenged by proprioceptive noise. In addition to the potential for testing this in developing organisms, our study has potential applications in soft robotics, as it is a way to determine the actuation pattern in complex situations where the best actuation pattern for a given goal may not be known {\it{a priori}}. 

\section*{Acknowledgments}
We thank Daniel Fortunato, Jordan Hoffmann and Vamsi Spandan for feedback. This work was supported in part by the grant W911NF-15-1-0166 from the US Army office of research.

  \setcounter{table}{0}
        \renewcommand{\thetable}{S\arabic{table}}%
        \setcounter{figure}{0}
        \renewcommand{\thefigure}{S\arabic{figure}}%
      \setcounter{equation}{0}
        \renewcommand{\theequation}{S.\arabic{equation}}
        \setcounter{section}{0}
        \renewcommand{\thesection}{S.\arabic{section}} 
\newpage
\section*{Appendix}

\subsection*{Parameter values}
\begin{table}[h]
\caption{Parameters and their values used in the simulation.}
\begin{tabular}{lll} \hline
 Symbol &Quantity & Value  \\ \hline \hline
 L & segment length & 1   \\
 $\tau_\theta$ & neuronal timescale & 1  \\
 $c\tau_\theta/k$ & scaled damping & 3.5   \\
 $f^m_{\max}/kL$ & scaled muscular force& 1   \\
 $\tau_m/\tau_\theta$ & scaled muscular timescale & 1   \\
 $f^f_{\max}/kL$ & scaled backward frictional force & 9   \\
 $\varepsilon^f$ & frictional smoothing & $10^{-6}$  \\
 $\eta$ & friction anisotropy & 30   \\
 $\Delta t$ & scaled discrete timestep & 0.01  \\ \hline
% $\alpha$ & learning rate & 0.001 &  & \\
% $\gamma$ & discount rate & 0.95 &  & \\
\end{tabular}
\label{tbl:nd-params}
\end{table}

\subsection*{Unlearned gait}
\begin{figure}[h]
\includegraphics[width=0.7\columnwidth]{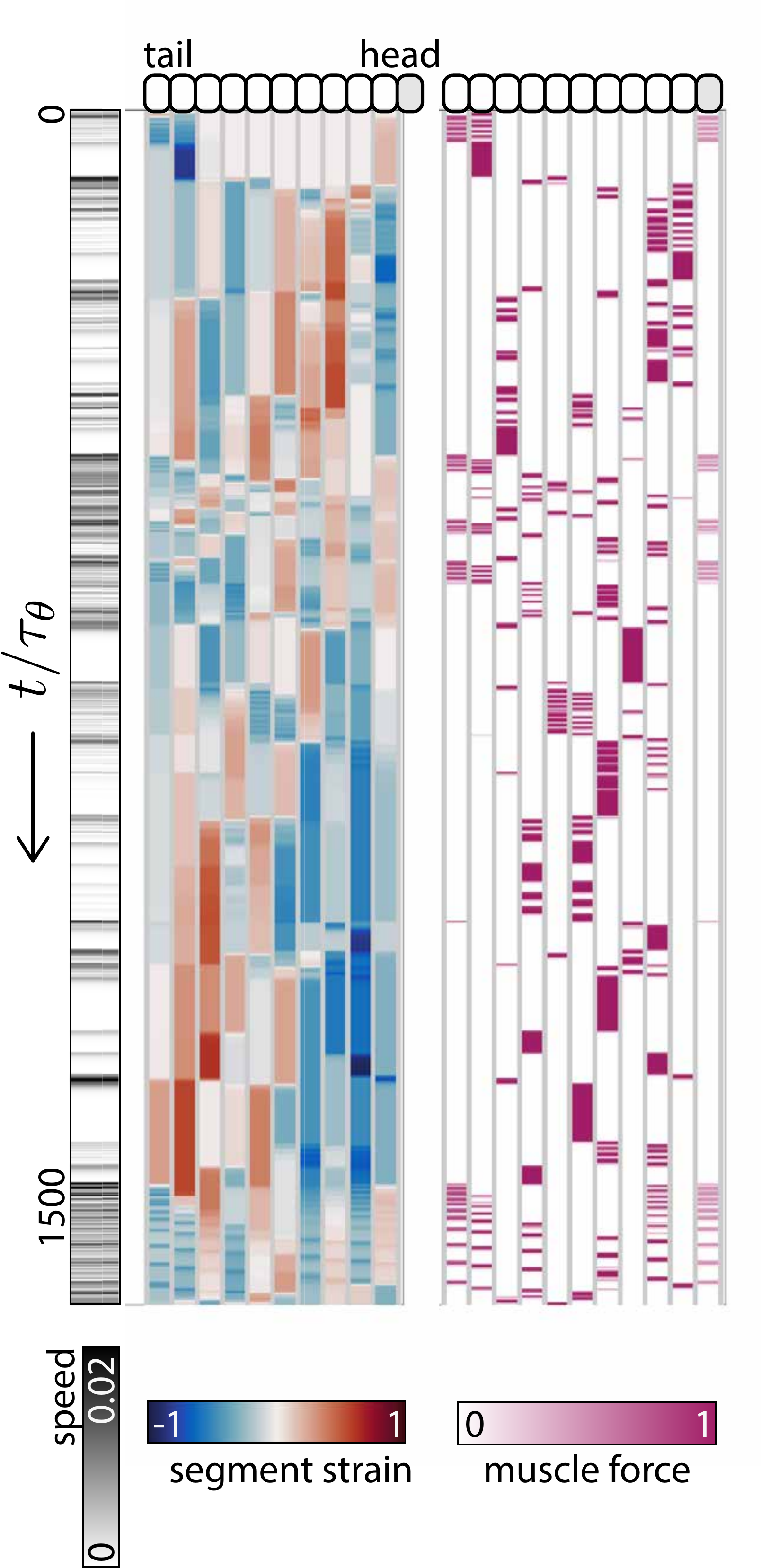}
\caption{Uncoordinated gait resulting from a fully connected neuronal network, as summarized by equations (1-8) of the main text. The speed of the centre of mass is in grey (left), corresponding segment strains in blue-red (middle) and muscle force in pink (right). The parameter values are summarized in Table S1. }
\label{fig:unlearned}
\end{figure} 

\subsection*{Videos} 
We include links to two videos that show the converged gait of the crawler with and without regularization, corresponding to Figures 2 A and B, respectively. 

\href{https://www.youtube.com/watch?v=_N195lCAdQ8&feature=youtu.be}{\color{blue} Regularized Gait}: Coordinated gait that arises from an initial uncoordinated gait with a regularization parameter $\epsilon = 0.01$.

\href{https://www.youtube.com/watch?v=eq8sWXoCP78&feature=youtu.be}{\color{blue} Unregularized Gait}: Coordinated gait that arises from an initial uncoordinated gait with no regularization parameter, i.e.\ $\epsilon = 0$, which leads to motion where multiple segments move concurrently. This gait is not robust to proprioceptive noise and is easily disrupted (see Figure 3C and corresponding text for details). 


\begin{thebibliography}{10}

\bibitem{chiel1997brain}
H.~J.~Chiel and R.~D.~Beer, "The brain has a body: adaptive behavior emerges from interactions of nervous system, body and environment." {\it Trends  Neurosci.} {\bf 20}, 553 (1997). 

\bibitem{rossignol2006dynamic}
S. Rossignol, R. Dubuc, and J.-P. Gossard, “Dynamic sensorimotor interactions in locomotion,” {\it Physiol. Rev.}, {\bf 86},  89–154, 2006.

\bibitem{pehlevan2016integrative}
C.~Pehlevan, P.~Paoletti, and L.~Mahadevan, "Integrative neuromechanics of crawling in D. melanogaster larvae," {\it Elife}, {\bf 5}, e11031 (2016). 

\bibitem{heckscher2012characterization}
E.~S.~Heckscher, S.~R.~Lockery, and C.~Q.~Doe, "Characterization of drosophila larval crawling at the level of organism, segment, and somatic body wall musculature," {\it J. Neurosci.} {\bf 32}, 12460 (2012). 

\bibitem{narayanan2014developmental}
D. Z. Narayanan and A. A. Ghazanfar, “Developmental neuroscience: How twitches make sense,” {\it Curr. Biol.}, {\bf 24}, pp. R971–R972, 2014.

\bibitem{berni2012autonomous}
J. Berni, S. R. Pulver, L. C. Griffith, and M. Bate, “Autonomous circuitry for substrate exploration in freely moving drosophila larvae,”  {\it Curr. Biol.} {\bf 22},  1861–1870, 2012.

\bibitem{eltringham1971life}
 S. K. Eltringham, {\it Life in mud and sand.} Crane, Russak, 1971.
 
 \bibitem{trueman1968burrowing}
 E. Trueman, “Burrowing habit and the early evolution of body cavities,” {\it Nature},  {\bf 218},  96, 1968.
 
\bibitem{suster2002embryonic}
M.~L.~Suster and M.~Bate, “Embryonic assembly of a central pattern generator without sensory input,” {\it Nature}, {\bf 416}, 174 (2002). 

\bibitem{Paoletti2014}
P.~Paoletti and L.~Mahadevan, “A proprioceptive neuromechanical theory of crawling,” {\it Proc. R. Soc.  B: Biol. Sci.} {\bf 281}, 20141092 (2014). 
 
\bibitem{sutton1998reinforcement}
R.~S.~Sutton, A.~G.~Barto, {\it Reinforcement learning: An introduction} (MIT press, 1998). 

\bibitem{silver2017mastering}
D.~Silver, J.~Schrittwieser, K.~Simonyan, I.~Antonoglou, A.~Huang, A.~Guez, T.~Hubert, L.~Baker, M.~Lai, A.~Bolton, {\it et al.} “Mastering the game of go without human knowledge,” Nature {\bf 550}, 354 (2017). 

\bibitem{colabrese2017flow}
S.~Colabrese, K.~Gustavsson, A.~Celani, and L.~Biferale, “Flow navigation by smart microswimmers via reinforcement learning,” {\it Phys. Rev. Lett.} {\bf 118}, 158004 (2017). 

\bibitem{reddy2016learning}
G.~Reddy, A.~Celani, T.~J.~Sejnowski, and M.~Vergassola, “Learning to soar in turbulent environments,” {\it Proc. Natl. Acad. Sci.} {\bf 113}, E4877 (2016). 

\bibitem{ermentrout1986}
G.~B.~Ermentrout and N.~Kopell, “Parabolic bursting in an excitable system coupled with a slow oscillation,” {\it SIAM J. App. Math.} {\bf 46}, 233 (1986)

\bibitem{hughes2007sensory}
C.~L.~Hughes and J.~B.~Thomas, A sensory feedback cir- cuit coordinates muscle activity in drosophila,” {\it Mol. Cell. Neuro.} {\bf 35}, 383 (2007).

\end{thebibliography}
\end{document}